\documentclass{article}
\pdfoutput=1
\usepackage{arxiv}

\usepackage[utf8]{inputenc} 
\usepackage[T1]{fontenc}    
\usepackage{hyperref}       
\usepackage{url}            
\usepackage{booktabs}       
\usepackage{amsfonts}       
\usepackage{nicefrac}       
\usepackage{microtype}      

\usepackage{graphicx}
\usepackage{booktabs}
\usepackage{float}
\usepackage{amsmath}
\usepackage{amssymb}
\usepackage{multirow}
\usepackage{adjustbox}
\usepackage{caption}
\usepackage{subcaption}
\usepackage{tikz, pgfplots}
\pgfplotsset{scaled y ticks=false}
\usepackage{footnote}
\usepackage{natbib}
\usepackage{doi}

\title{Introducing Three New Benchmark Datasets for Hierarchical Text Classification}

\date{} 					

\author{
	\href{https://orcid.org/0000-0001-6782-2381}{\includegraphics[scale=0.06]{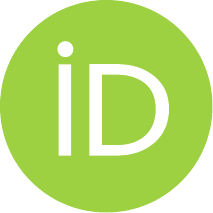}\hspace{1mm}Jaco du Toit} \\
	Computer Science Division, Department of Mathematical Sciences\\
	Stellenbosch University\\
	Stellenbosch, South Africa \\
	\texttt{jacowdutoit11@gmail.com} \\
	\And
	\href{https://orcid.org/0000-0001-6099-3057}{\includegraphics[scale=0.06]{orcid.pdf}\hspace{1mm}Herman Redelinghuys} \\
	Centre for Research on Evaluation, Science and Technology\\
	Stellenbosch University\\
	Stellenbosch, South Africa \\
	\texttt{hredelinghuys@sun.ac.za} \\
	\And
	\href{https://orcid.org/0000-0003-1957-3979}{\includegraphics[scale=0.06]{orcid.pdf}\hspace{1mm}Marcel Dunaiski} \\
	Computer Science Division, Department of Mathematical Sciences\\
	Stellenbosch University\\
	Stellenbosch, South Africa \\
	\texttt{marceldunaiski@sun.ac.za} \\
}

\renewcommand{\shorttitle}{Introducing Three HTC Benchmark Datasets}

\hypersetup{
pdftitle={Introducing Three New Benchmark Datasets for Hierarchical Text Classification},
pdfauthor={Jaco du Toit, Herman Redelinghuys, Marcel Dunaiski},
pdfkeywords={Document Classification, Hierarchical Text Classification, Benchmark Datasets, Large Language Models},
}

\begin{document}
\maketitle

\begin{abstract}
Hierarchical Text Classification (HTC) is a natural language processing task with the objective to classify text documents into a set of classes from a structured class hierarchy. Many HTC approaches have been proposed which attempt to leverage the class hierarchy information in various ways to improve classification performance. Machine learning-based classification approaches require large amounts of training data and are most-commonly compared through three established benchmark datasets, which include the Web Of Science (WOS), Reuters Corpus Volume 1 Version 2 (RCV1-V2) and New York Times (NYT) datasets. However, apart from the RCV1-V2 dataset which is well-documented, these datasets are not accompanied with detailed description methodologies. In this paper, we introduce three new HTC benchmark datasets in the domain of research publications which comprise the titles and abstracts of papers from the Web of Science publication database. We first create two baseline datasets which use existing journal-and citation-based classification schemas. Due to the respective shortcomings of these two existing schemas, we propose an approach which combines their classifications to improve the reliability and robustness of the dataset. We evaluate the three created datasets with a clustering-based analysis and show that our proposed approach results in a higher quality dataset where documents that belong to the same class are semantically more similar compared to the other datasets. Finally, we provide the classification performance of four state-of-the-art HTC approaches on these three new datasets to provide baselines for future studies on machine learning-based techniques for scientific publication classification.
\end{abstract}


\section{Introduction}
Hierarchical text classification (HTC) approaches are used to categorise text documents into a set of classes from a hierarchical class structure based on the textual content of the documents. HTC approaches are well-suited for the organisation of large document collections since they enable users to select the level of granularity that they prefer based on the class hierarchy such that they can reduce their search scope to a smaller subset of documents. 

Due to the far-reaching applications of HTC, many approaches have been proposed in recent years. These approaches aim to leverage the hierarchical class structure through different techniques in order to improve classification performance~\citep{Chen,Deng,dutoit,GopalYang,Huang2,Jiang,Mao,Peng,Wang1,Wang2,Wu,Zhou}.

These text-based HTC approaches are most-commonly evaluated through three established benchmark datasets, namely the Web Of Science (WOS)~\citep{Kowsari}, Reuters Corpus Volume 1 Version 2 (RCV1-V2)~\citep{RCV1} and the New York Times (NYT) Annotated Corpus~\citep{nyt}. However, only the creators of the RCV1-V2 dataset provide a detailed methodology for the creation of their dataset~\citep{RCV1}. In particular, the WOS dataset, which is the only benchmark HTC dataset in the domain of research publications, does not provide sufficient detail on how the dataset was created.

We believe it is important to provide a more detailed description of the dataset creation methodology to facilitate reproducibility and reliable comparisons between different classification approaches. Furthermore, detailed dataset creation methodologies enable a better analysis of the results obtained by classification approaches since the characteristics of the dataset may influence the performance of different approaches. 

In this paper, we propose three new datasets for HTC tasks in the domain of research publications. As a starting point for developing the new datasets, we use data that comprises the title and abstract of academic publications from the Web of Science publication database. Our three datasets each use this data but have different classification schemas which determine the categories assigned to the publications.

First, we use a journal-based classification schema from Web of Science which assigns categories to each journal and classifies a publication based on the journal it is published in. However, journal-based classifications have been shown to be unreliable and often inaccurate, with \citet{Shu2019ComparingJournal} showing that these classification schemas may incorrectly classify almost half of publications in some cases. \citet{Wang2016LargeScale} compare the two most popular journal-based classification schemas (the WOS subject categories and the Scopus subject areas) and show that they often assign the same categories to publications which do not have strong citation-relationships between them. This indicates that the classifications are too lenient. Furthermore, journal-based classification schemas are not well suited to multidisciplinary journals such as Nature, Science, and PNAS for the obvious reason that these journals span multiple distinct disciplines which leads to publications being incorrectly assigned to all of the categories in the journal. \citeauthor{Wang2016LargeScale} state that the popularisation of open access multidisciplinary journals such as PLoS ONE and Scientific Reports further increase the unreliability of journal-based classifications.

Our second proposed dataset uses a citation-based classification schema recently developed by Clarivate~\citep{traag2019leiden}. Citation-based classification schemas use the citation relationships from a collection of research publications to form clusters of documents which share the same class. However, the citation-based classification schema proposed by Clarivate does not allow a document to belong to multiple research fields, which prevents multi-disciplinary research publications from being appropriately classified. 

Due to the shortcomings of these two classification schemas, we also propose an approach that combines these two schemas to create a new categorisation that leverages their respective advantages. In our approach, we filter out categories and documents which do not have a clear overlap between the journal-and citation-based classifications. Therefore, we create new class assignments which are formed by removing assignments and categories that do not form obvious mappings between the two existing classification schemas. The objective of this approach is to increase the probability that an individual document is correctly classified since its classifications are assigned based on its content (journal) as well as its position in the citation network. Furthermore, our proposed approach allows documents to belong to multiple classes, which is important for multi-disciplinary publications, while leveraging the finer-grained citation-based classifications to improve the category assignments of documents.

In summary, we create three new datasets of which the first two use existing journal-and citation-based classifications respectively, while the third dataset uses the classifications obtained by combining these two classifications and applying our proposed filtering approach. Our datasets are unique among benchmark HTC datasets since we sample documents equally for each of the classes in the second level of the hierarchy. Therefore, our datasets are significantly more balanced than the currently used benchmark HTC datasets and therefore better suited for machine learning-based classification approaches.

To evaluate the quality of the three datasets we analyse the semantic similarity between the documents belonging to the different classes. This is done by encoding the documents for each class into a semantic vector embedding and measuring the distances between the documents in the embedding space. We show that the new dataset based on the combination of the journal- and citation-based classification improves the average similarity of the documents belonging to a specific class as well as the separation between documents in different classes.

Finally, we perform classification experiments on the three proposed datasets with state-of-the-art HTC approaches. These experiments provide further insights to the capabilities the classification approaches as well as the difficulty in accurately classifying the publications in the three proposed datasets.

\renewcommand{\arraystretch}{1.5}
\begin{table}[th!]
    \caption[Example publications with associated WOS subject categories.]{Example publications with associated WOS subject categories. The ``Publication'' column comprises the title and truncated abstract of the publication.}
    \label{tab:WOSCDatasetEx}
    \small
    \centering
    \begin{tabular*}{\textwidth}{p{0.7\textwidth}p{0.35\textwidth}}
        \toprule
        \multicolumn{1}{l}{\normalsize Publication} & \multicolumn{1}{l}{\normalsize WOS categories} \\
        \midrule
         ``Can Creditor Bail-in Trigger Contagion? The Experience of an Emerging Market. The successful bail-in of creditors in African Bank, a small South African monoline lender, provides an opportunity to evaluate the intended and unintended consequences of new resolution tools. Using a dataset that matches quarterly, daily, and financial-instrument level data, I show that the bail-in led to money-market funds `breaking the buck,' triggering significant redemptions and some financial contagion...'' & $\bullet$ Business \newline $\bullet$ Finance \newline $\bullet$ Economics \\
         ``Dissecting the genre of Nigerian music with machine learning models. Music Information Retrieval (MIR) is the task of extracting high-level information, such as genre, artist or instrumentation from music. Genre classification is an important and rapidly evolving research area of MIR. To date, only a small amount of research work has been done on the automatic genre classification of Nigerian songs. Hence, this study presents a new music dataset, namely the ORIN dataset, consisting of only Nigerian songs...'' & $\bullet$ Computer Science \newline $\bullet$ Information Systems \\
         ``The complementarity of a diverse range of deep learning features extracted from video content for video recommendation. Following the popularisation of media streaming, a number of video streaming services are continuously buying new video content to mine the potential profit from them. As such, the newly added content has to be handled well to be recommended to suitable users. In this paper, we address the new item cold-start problem by exploring...'' & $\bullet$ Computer Science \newline $\bullet$  Artificial Intelligence \newline $\bullet$ Engineering (Electrical \& \newline~~Electronic) \newline $\bullet$ Operations Research \& \newline~~Management Science \\
    \bottomrule
    \end{tabular*}
\end{table}
\renewcommand{\arraystretch}{1.0}

\section{Background}

\subsection{Hierarchical Text Classification}
HTC tasks model class hierarchies as a directed acyclic graph $\mathcal{H} = (C, E)$, where $C = {c_1, \ldots, c_L}$ represents the set of class nodes, with $L$ denoting the total number of classes, and $E$ representing the edges that establish the hierarchical relationships between these nodes. For the HTC tasks that we consider, each node, except the root, is connected to a single parent, resulting in a tree structure. The objective of HTC is to classify a input sequence (i.e., text document) composed of $T$ tokens $\mathbf{x} = [x_1, \ldots, x_T]$ into a subset of classes $Y’ \subseteq C$, which maps to one or more paths within the hierarchy $\mathcal{H}$. 

Many HTC approaches have been proposed in recent years, and these approaches incorporate the hierarchical class structure information into the classification process through various ways in order to improve performance~\citep{Chen,Deng,dutoit,GopalYang,Huang2,Jiang,Mao,Peng,Wang1,Wang2,Wu,Zhou}. In particular, the most recently proposed HTC approaches combine the hierarchical class structure with the natural language understanding capabilities of large language models (LLM)~\citep{electra,Devlin,roberta} in order to obtain state-of-the-art results on benchmark datasets.

\citet{Wang2} introduced the Hierarchy-aware Prompt Tuning (HPT) method, which enhances the input sequence fed into the LLM by incorporating hierarchy-aware prompts derived from a graph encoder. These prompts, along with a masked token for each level of the class hierarchy, transform the HTC task to more closely resemble the MLM pre-training task~\citep{Vaswani}. More recently, \citet{dutoit2} also proposed the Hierarchy-aware Prompt Tuning for Discriminative PLMs (HPTD) approach which which extends the HPT approach to discriminative language models~\citep{electra,He2021DeBERTaV3}. They investigated the use of different discriminative language models such as ELECTRA~\citep{electra} and DeBERTaV3~\citep{He2021DeBERTaV3} such that their models are referred to as HPTD-ELECTRA and HPTD-DeBERTaV3 respectively. Furthermore, \citet{dutoit} proposed the the global hierarchical label-wise attention (GHLA) approach which uses label-wise attention mechanisms to dynamically place more attention on the most relevant features for each class separately. They investigated different base LLM architectures and found that using RoBERTa~\citep{roberta} obtained the highest classification performance. Their best-performing model is therefore referred to as GHLA$_\text{RoBERTa}$ in this paper.

\subsection{Web of Science Subject Categories}
\label{Section:OGWOSclassifcation}
The WOS subject categories form a classification schema which categorises research publications based on the journal, conference, or book in which they are published. For sake of brevity, we refer to all research publication venues as journals for the remainder of this paper.

In the WOS subject categories classification schema, each journal is assigned to one or more categories from 256 possible classes which cover all thematic areas of scientific research. Therefore, a document is assigned to the classes associated with the journal it is published in. The WOS subject categories are not structured hierarchically, i.e., they constitute a multi-label classification framework. Table~\ref{tab:WOSCDatasetEx} shows the title and abstract of example publications with their assigned WOS categories. 

The detailed methodology for assigning WOS categories to a journal is not published, but these classifications consider many aspects which include: the content and scope of the journal, affiliations of authors and editors of the journal, citation relationships of publications in the journal~\citep{Pudovkin2002AlgorithmicProcedure}, funding agencies and sponsors, as well as the journal's categorisation in other bibliographic databases.

\subsection{CREST Journal-based Classification Schema}
\label{Section:CRESTclassifcation}
The Centre for Research on Evaluation, Science and Technology (CREST) is a research centre in the Faculty of Arts and Social Sciences at Stellenbosch University which focuses on various topics including bibliometrics, scientometrics, and research evaluation. CREST devised a new classification schema with a two-level hierarchical class structure as an alternative to the WOS subject categories. 

They mapped the WOS categories to a two-level hierarchical class structure with fewer classes to balance the size of class clusters and cover strategic fields of research. Their schema comprises 7 and 55 classes for the first and second levels of the class hierarchy respectively. Table~\ref{tab:CRESTMappingEx} provides examples of the mappings from WOS categories to the classification schema proposed by CREST which we refer to as Journal-based Topics (JT), where JT$_{\text{L}1}$ and JT$_{\text{L}2}$ represent the classes for level 1 and level 2 respectively. Table~\ref{tab:CRESTDatasetEx} shows example publications with their JT classifications.

\begin{table}[h]
    \caption{Examples of mapping from WOS categories to JT.}
    \small
    \centering
    \begin{tabular*}{\textwidth}{p{0.33\textwidth}p{0.33\textwidth}p{0.33\textwidth}}
        \toprule
        WOS & JT$_{\text{L}1}$ & JT$_{\text{L}2}$ \\
        \midrule
        Agronomy & Agricultural sciences & Agronomy \\
        Forestry & Agricultural sciences & Agricultural sciences (Other) \\
        Literature & Humanities and arts & Language \& linguistics \\
        Art & Humanities and arts & Other humanities \& arts \\
        Geology & Natural sciences & Geosciences \\
        Behavioral Sciences & Social sciences & Psychology \\
        \bottomrule
    \end{tabular*}
  \label{tab:CRESTMappingEx}
\end{table}

\renewcommand{\arraystretch}{1.5}
\begin{table}[h]
    \caption[Example publications with associated JT classifications.]{Example publications with associated JT classifications. The ``Publication'' column comprises the title and truncated abstract of the publication.}
    \small
    \label{tab:CRESTDatasetEx}
    \begin{tabular*}{\textwidth}{p{0.4\textwidth}p{0.2\textwidth}p{0.4\textwidth}}
        \toprule
        \multicolumn{1}{l}{\normalsize Publication} & \multicolumn{1}{l}{\normalsize JT$_{\text{L}1}$} & \multicolumn{1}{l}{\normalsize JT$_{\text{L}2}$} \\
        \midrule
        ``Can Creditor Bail-in Trigger Contagion? The Experience of an Emerging Market...'' & $\bullet$ Social Sciences & $\bullet$ Business \newline $\bullet$ Economics\\ 
        ``Dissecting the genre of Nigerian music with machine learning models. Music Information...'' & $\bullet$ Natural sciences & $\bullet$ Information, computer \& \newline~~communication technologies \\
        ``The complementarity of a diverse range of deep learning features extracted from video content for video recommendation. Following the popularisation of media streaming, a number of video streaming services are...'' & $\bullet$ Engineering \newline $\bullet$ Natural sciences & $\bullet$ Electrical \& electronic engineering \newline $\bullet$ Engineering sciences (other) \newline $\bullet$ Information, computer \&\newline~~communication technologies \\
        \bottomrule
    \end{tabular*}
\end{table}
\renewcommand{\arraystretch}{1.0}

\subsection{Clarivate Citation Topics Classification Schema}
Clarivate proposed the Citation Topics (CT) classification schema, which clusters documents based on citation relationships, as an alternative to journal-based classifications such as the WOS subject categories. Although the detailed methodology for the classification algorithm has not been published, the classifications are obtained by a Leiden community detection algorithm~\citep{traag2019leiden} based on the citation relationships between documents. Community detection algorithms have the objective of identifying communities (or clusters) of nodes within a network that are more densely connected to each other than to nodes outside of the community~\citep{traag2019leiden}. These algorithms typically use an objective function which measures the quality of the node partitions with the aim of maximising the intra-community connections while minimising the inter-community connections. The Leiden community detection algorithm proposes several improvements over the popular Louvain community detection algorithm~\citep{blondel2008louvain} to obtain a better partition of nodes in a network. 

The Louvain algorithm starts with each node forming its own community and attempts to optimise the modularity of the network which is calculated as:
\begin{equation}
    \mathcal{M} = \frac{1}{2m}\sum_{i=c}^C(e_c-\gamma\frac{K_{c}^2}{2m})
\end{equation}
\noindent where $m$ is the total number of edges in the network, $C$ is the number of communities, and $K_c$ is the sum of the degrees of the nodes in community $c$ such that $\frac{K_{c}^2}{2m}$ represents the expected number of edges in the community. Furthermore $\gamma$ is a resolution parameter which determines the number of communities formed by the algorithm since higher $\gamma$ values lead to more communities being formed. Therefore, the objective of the algorithm is to maximise the difference between the expected and true number of edges in a community. Note that the Louvain algorithm can be optimised through other objective functions~\citep{traag2019leiden}. 

\renewcommand{\arraystretch}{1.5}
\begin{table}[th]
    \caption[Example publications with associated citation-based classifications.]{Example publications with associated citation-based classifications at hierarchical levels L1 through L3. The ``Publication'' column in this example comprises the title and truncated abstract of the publication.}
    \small
    \label{tab:CTDatasetEx}
    \begin{tabular*}{\textwidth}{p{0.35\textwidth}p{0.20\textwidth}p{0.17\textwidth}p{0.15\textwidth}}
    \toprule
    \multicolumn{1}{l}{\normalsize Publication} & \multicolumn{1}{l}{\normalsize CT$_{\text{L}1}$} & \multicolumn{1}{l}{\normalsize CT$_{\text{L}2}$} & \multicolumn{1}{l}{\normalsize CT$_{\text{L}3}$} \\
    \midrule
    ``Can Creditor Bail-in Trigger Contagion? The Experience of an Emerging...'' & $\bullet$ Social Sciences & $\bullet$ Economics  & $\bullet$ Economic \newline Growth \\
    ``Dissecting the genre of Nigerian music with machine learning models. Music Information...'' & $\bullet$ Electrical Engineering, \newline Electronics \& \newline Computer Science & $\bullet$ Knowledge \newline Engineering \& \newline Representation & $\bullet$ Statistical Tests \\
    ``The complementarity of a diverse range of deep learning features extracted from video content for video...'' & $\bullet$ Electrical Engineering, \newline Electronics \& \newline Computer Science & $\bullet$ Knowledge \newline Engineering \& \newline Representation & $\bullet$ Collaborative \newline Filtering \\
    \bottomrule
    \end{tabular*}
\end{table}
\renewcommand{\arraystretch}{1.0}

The Louvain algorithm uses two phases to assign nodes to suitable communities. In the first phase, each node is considered to be moved to neighbouring communities and the change in objective function is determined. The node movements that maximise the increase in the objective function are performed until no improvement in the objective function can be achieved. In the second phase, the nodes in the communities obtained by the first phase are aggregated to form individual nodes where the edge weights between the nodes are determined by the sum of the weights of the edges between the original nodes in the communities. The first phase is applied to the aggregated nodes and this procedure is repeated until the objective function can no longer be improved. \citet{traag2019leiden} show that the Louvain algorithm may obtain arbitrarily badly connected communities and even internally disconnected communities, i.e., communities where a section of the community can only reach another section of the community through a path that goes outside of that community. 

The Leiden algorithm addresses these shortcomings by proposing several improvements to the Louvain algorithm. Similar to the Louvain algorithm, each node starts off as its own community and the nodes are moved to different communities to maximise an objective function. However, the Leiden algorithm introduces an approach which refines these communities obtained by the first phase such that each community may be split into multiple sub-communities. The refinement phase uses the partitions obtained by the first phase to locally merge the nodes in each community to potentially form sub-communities. In the refinement phase, nodes are not always greedily assigned to the community which maximises the objective function increase. Alternatively, any node assignment which results in an increase in the objective function is considered. The assignments are selected with a certain probability based on their increase in the objective function such that assignments with larger increases are more likely. This non-greedy merging in the refinement phase enables a better exploration of the partition space~\citep{traag2019leiden}. The aggregate network is created based on the refined partitions which increases the probability of finding high-quality communities. Using these improvements, the Leiden algorithm guarantees that formed communities are well connected and that it converges to a solution where all subsets of communities are guaranteed to be locally optimal.

In the CT classification schema, the document communities (or clusters) obtained by the Leiden algorithm are used to assign documents to a three-level hierarchical class structure. The first and second level classes (CT$_{\text{L}1}$ and CT$_{\text{L}2}$) are manually labelled based on the contents of the documents in their clusters, while the third level classes (CT$_{\text{L}3}$) are labelled algorithmically with the most significant keyword in the cluster. The three levels of the class hierarchy comprise 10, 326, and 2\,457 classes respectively. Table~\ref{tab:CTDatasetEx} shows example publications with their CT classifications.

\section{Methodology}
From the WOS publication database, we randomly sampled 5\,000 papers for each of the level two citation-based classes (CT$_{\text{L}2}$), resulting in 1\,630\,000 records. Each record contains the title and abstract of the publication, along with the journal-based (JT) and citation-based (CT) classifications. We use this dataset to create three HTC datasets which include:
\begin{itemize}
    \item WOS$_\text{JT}$, which only uses the journal-based JT classifications.
    \item WOS$_\text{CT}$, which only uses the citation-based CT classifications.
    \item WOS$_\text{JTF}$, which uses the JT Filtered (JTF) classification schema that filters out documents and classes which do not have a clear overlap between JT and CT classifications.
\end{itemize}

Table~\ref{tab:DatasetsSummary} shows the summary statistics for the three datasets. We split each dataset into train (70\%), development (15\%), and test (15\%) sets.

\begin{table}[h]
 \caption[Characteristics of the newly created HTC datasets.]{Characteristics of the newly created HTC datasets. The column ``Levels'' gives the number of levels, while ``Classes$_{\text{L}1}$'' and ``Classes$_{\text{L}2}$'' give the number of first- and second-level classes in the class structure. ``Avg.~Classes'' is the average number of classes per document, while ``Train'', ``Dev'',  and ``Test'' are the number of instances in each of the dataset splits.}
 \label{tab:DatasetsSummary}
 \centering
 \begin{tabular*}{\textwidth}{cccccccc}
    \toprule
    \normalsize Dataset & \normalsize Levels & \normalsize Classes$_{\text{L}1}$ & \normalsize Classes$_{\text{L}2}$ & \normalsize Avg.~Classes & \normalsize Train & \normalsize Dev & \normalsize Test \\
    \midrule
    WOS$_\text{JT}$ & 2 & 6 & 52 & 2.93 & 30\,356 & 6\,505 & 6\,505 \\
    WOS$_\text{CT}$ & 2 & 10 & 326 & 2.00 & 45\,640 & 9\,780 & 9\,780 \\
    WOS$_\text{JTF}$ & 2 & 6 & 46 & 2.25 & 30\,048 & 6\,439 & 6\,439 \\
    \bottomrule
    \end{tabular*}
\end{table}

\subsection{WOS$_\text{JT}$ Dataset}
To create the WOS$_\text{JT}$ dataset we use the JT classifications as described above but remove the JT classes for which there are no instances in the sampled dataset. We randomly sample 1\,000 documents for each JT$_{\text{L}2}$ class with the aim of creating a dataset that is balanced at the second level of the class hierarchy. Since a document can be assigned to one or more JT$_{\text{L}2}$ classes, the second level classes are not perfectly balanced and the dataset contains 43\,366 documents in total. 

\begin{figure}[ht]
    \centerline{\begin{tikzpicture}[scale = 0.65, font = \Large]
  \begin{axis}[
    symbolic x coords = {Agricultural Sciences, Humanities and Arts, Engineering, Social Sciences, Health Sciences, Natural Sciences},
    axis y line*=left,
    ybar,
    ylabel={Number of Documents},
    ymin = 4000, ymax = 16000,
    xlabel near ticks,
    xticklabel style={rotate=45, anchor=east, font=\large},
    xtick = data,
    ylabel near ticks,
    bar width=1.3cm,
    enlarge x limits= 0.2,
    nodes near coords align={center},
    width=18cm,
    height=10cm,
    legend style={
      at={(0.41,0.89)},
      draw=none
    },
  ]
\addplot[draw = black, fill = cyan, line width=1pt] 
    coordinates {(Agricultural Sciences, 4883) (Humanities and Arts, 6151) (Engineering, 7385) (Social Sciences, 10767) (Health Sciences, 10954) (Natural Sciences, 15592)};
\legend{Number of Documents}
\end{axis}
  
  \begin{axis}[
    symbolic x coords = {Agricultural Sciences, Humanities and Arts, Engineering, Social Sciences, Health Sciences, Natural Sciences},
    ymin = 0, ymax = 15,
    hide x axis,
    axis y line*=right,
    yticklabel pos=right,
    ylabel={Number of Children},
    ylabel near ticks,
    nodes near coords align={center},
    enlarge x limits=0.2,
    width=18cm,
    height=10cm,
    legend style={
      at={(0.385,0.97)},
      draw=none
    },
  ]
      \addplot[color = black, mark = square*, line width=1pt, mark size=3pt] coordinates {(Agricultural Sciences, 5) (Humanities and Arts, 6) (Engineering, 7) (Social Sciences, 10) (Health Sciences, 10) (Natural Sciences, 14)};
      \legend{Number of Children}
  \end{axis}
  \draw[draw=black] (0.25,6.72) rectangle (6.7,8.15);
  
\end{tikzpicture}}
\caption[Characteristics of first-level classes of the WOS$_\text{JT}$ dataset.]{Characteristics of first-level classes of the WOS$_\text{JT}$ dataset. The bar plot gives the number of documents assigned to each class (left y-axis) while the line plot shows the number of children for each class (right y-axis).}
\label{fig:CREST_stats}
\end{figure}
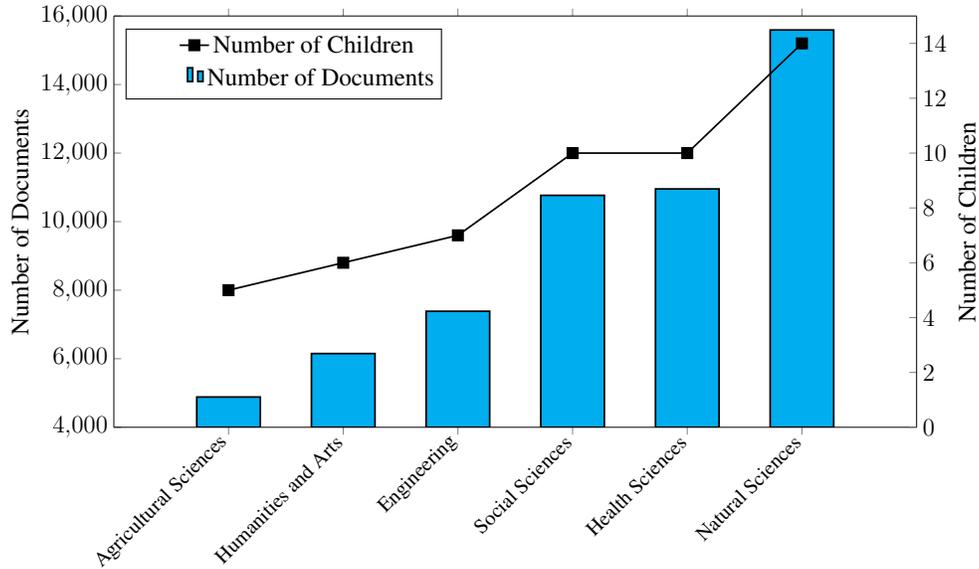

Figure~\ref{fig:CREST_stats} presents the first-level classes of the resulting WOS$_\text{JT}$ dataset along with the associated number of children classes and the number of documents assigned to each class. This figure shows how the number of documents assigned to a first-level class is larger for those classes with a larger number of children classes due to our sampling strategy. 

Figure~\ref{fig:CREST_overlap_level_1} presents the co-occurrence document counts for the first-level classes, i.e., the number of times that a document with a certain class is also assigned to each of the other classes. This figure shows that categories such as ``Natural sciences'' and ``Engineering'' are much more likely to co-occur than other category combinations, for example, ``Humanities and arts'' and ``Agricultural sciences''. 

\begin{figure}[h]
  \centerline{\scalebox{1}
  {\begin{tikzpicture}[scale = 0.65, font = \Large]
        \begin{axis}[
            ybar,
            xticklabel=\pgfmathprintnumber\tick,
            xlabel={Documents per Class},
            ylabel={Frequency},
            ymin=0,
            bar width=0.2cm,
            enlarge x limits=0.02,
            nodes near coords align={center},
            width=20cm,
            height=10cm,
            xmin=500,
            xmax=4000,
            xtick={500,1000,1500,2000,2500,3000,3500,4000},
            xticklabels={500,1000,1500,2000,2500,3000,3500,4000},
        ]
        
        \addplot+[hist={bins=28}, fill=cyan, draw=black, line width=1pt]
        table[y index=0] {CREST_lower_class_count.txt};
        \end{axis}

        \begin{axis}[
        hide axis,
        at={(0,0)},
        width=20cm,
        height=10cm,
        xmin=500,
        xmax=4000,
        enlarge x limits=0.02,
    ]
    
    \addplot[color=blue, mark=none, line width=1.2pt] table {CREST_kde_data.txt};
    
    \end{axis}
\end{tikzpicture}}}
    \caption{Distribution of the documents assigned to each of the JT$_{\text{L}2}$ classes.}
\label{fig:CREST_label_counts_level_2}
\end{figure}
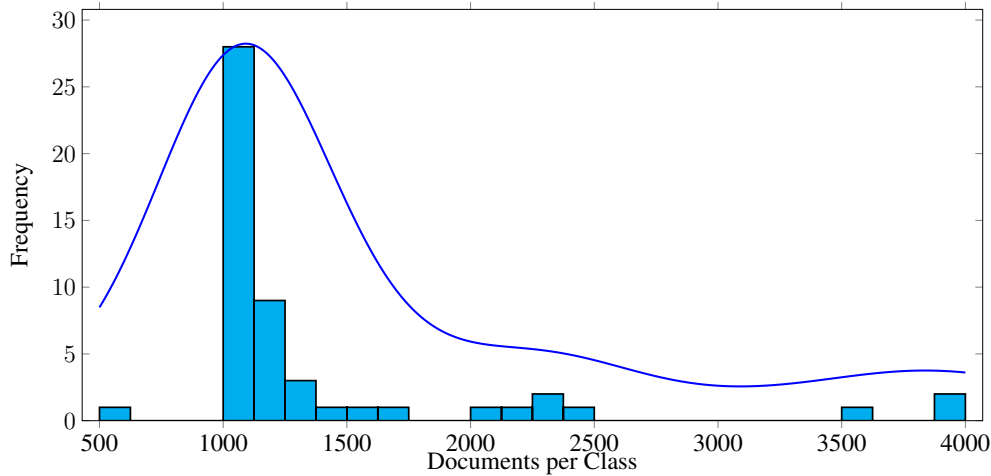

Figure~\ref{fig:CREST_label_counts_level_2} presents the distribution of the number of documents assigned to each of the classes in the second level of the hierarchy. From this figure we can see that most JT$_{\text{L}2}$ classes have around 1\,000 document assignments. However, the ``Ornithology'' class has less than 1\,000 document assignments since it only has 588 documents available to sample from. Furthermore, the ``Other social sciences'', ``Clinical and public health (other)'', and ``Engineering sciences (other)'' classes have 3\,553, 3\,900, and 3\,984 documents respectively since they often overlap with other class assignments.

\begin{figure}[h]
  \centerline{\scalebox{1.1}{\begin{tikzpicture}

    \definecolor{lightred}{RGB}{255, 50, 0}
    
    \pgfplotsset{
        colormap={mypurpleyellow}{
        	color(0cm) = (yellow);
        	color(6cm) = (red);
    	}
	}
    \def\matrixdata{{4883,109,775,6,1322,149},
{109,7385,307,40,3301,702},
{775,307,10954,280,2134,1261},
{6,40,280,6151,303,1208},
{1322,3301,2134,303,15592,1423},
{149,702,1261,1208,1423,10767}}
    
    \def\rowlabels{{"Agricultural Sciences", "Engineering", "Health Sciences", "Humanities and Arts", "Natural Sciences", "Social Sciences"}}
    
    \foreach \row [count=\y] in \matrixdata {
        \foreach \cell [count=\x] in \row {
            \pgfmathsetmacro\normalizedvalue{100*(ln(\cell+1)/ln(16000))}
            \node[fill=lightred!\normalizedvalue!yellow, text=black, minimum size=1cm, inner sep=0pt, anchor=center, font=\scriptsize] at (\x, -\y) {\cell};
        }
    }

    \foreach \y/\label in {1/Agricultural Sciences, 2/Engineering, 3/Health Sciences, 4/Humanities and Arts, 5/Natural Sciences, 6/Social Sciences} {
        \node[anchor=east, font=\scriptsize] at (0.45, -\y) {\label};
        \draw (0.4, -\y) -- (0.5, -\y);
    }

    \foreach \x/\label in {1/Agricultural Sciences, 2/Engineering, 3/Health Sciences, 4/Humanities and Arts, 5/Natural Sciences, 6/Social Sciences} {
        \node[anchor=east, font=\scriptsize, rotate=45] at (\x+0.05, -6.6) {\label};
        \draw (\x, -6.5) -- (\x, -6.6);
    }

    \draw[black, thin] (0.5, -0.5) rectangle (6.5, -6.5);

    \begin{axis}[
        at={(800,-130)},  
        anchor=south west,
        scale only axis,
        width=1cm,
        height=6cm,  
        colorbar,  
        hide axis,  
        point meta min=0,
        point meta max=16000,  
        colorbar style={
            ytick={0, 16000},  
            yticklabel style={font=\tiny},
            yticklabel pos=right,  
            colormap name=mypurpleyellow,  
            ylabel={Document Count},
            scaled ticks=false,
            ylabel style={font=\scriptsize, at={(2, 0.27)}, anchor=west},
            font=\scriptsize
        },
    ]
    \end{axis}
    
\end{tikzpicture}}}
    \caption{Document co-occurrence counts for first-level classes of the WOS$_\text{JT}$ dataset.}
\label{fig:CREST_overlap_level_1}
\end{figure}
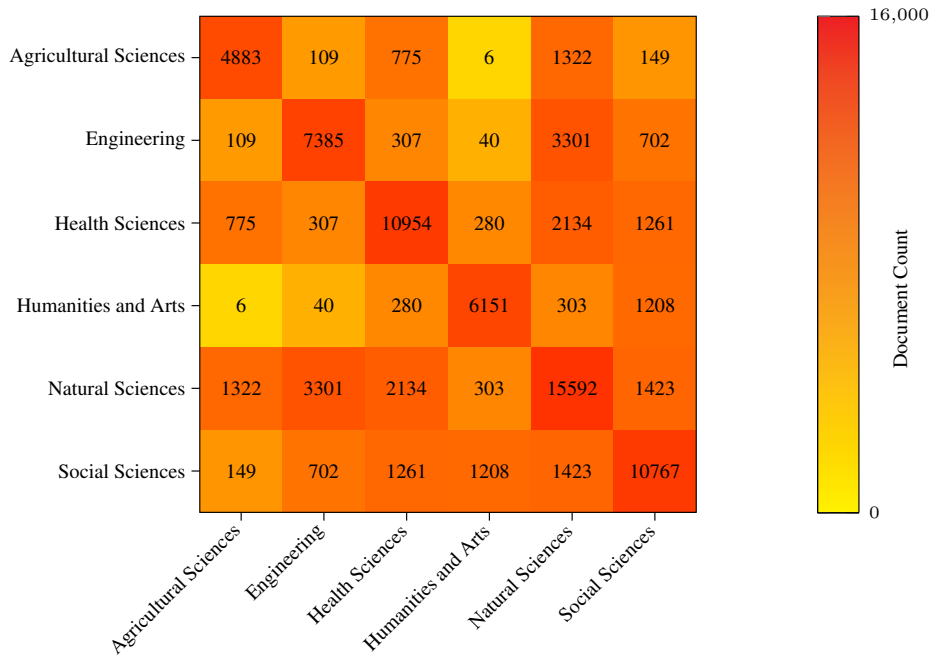

\subsection{WOS$_\text{CT}$ Dataset}
We create the WOS$_\text{CT}$ dataset by only using the citation-based CT classifications as described above. However, we only use the first two levels of the class hierarchy since the third level consists of 2\,457 classes which leads to a severely imbalanced dataset with certain classes at the third level having a much higher number of documents than others. Therefore, WOS$_\text{CT}$ uses 10 first-level (CT$_{\text{L}1}$) and 326 second-level (CT$_{\text{L}2}$) classes. We randomly sample 200 documents for each CT$_{\text{L}2}$ class to create the WOS$_\text{CT}$ dataset with 65\,200 academic publications. Since each document is assigned a single class per level, there are no co-occurrences between the classes at the same level, resulting in a perfectly balanced dataset at the second level with exactly 200 documents per CT$_{\text{L}2}$ class. 

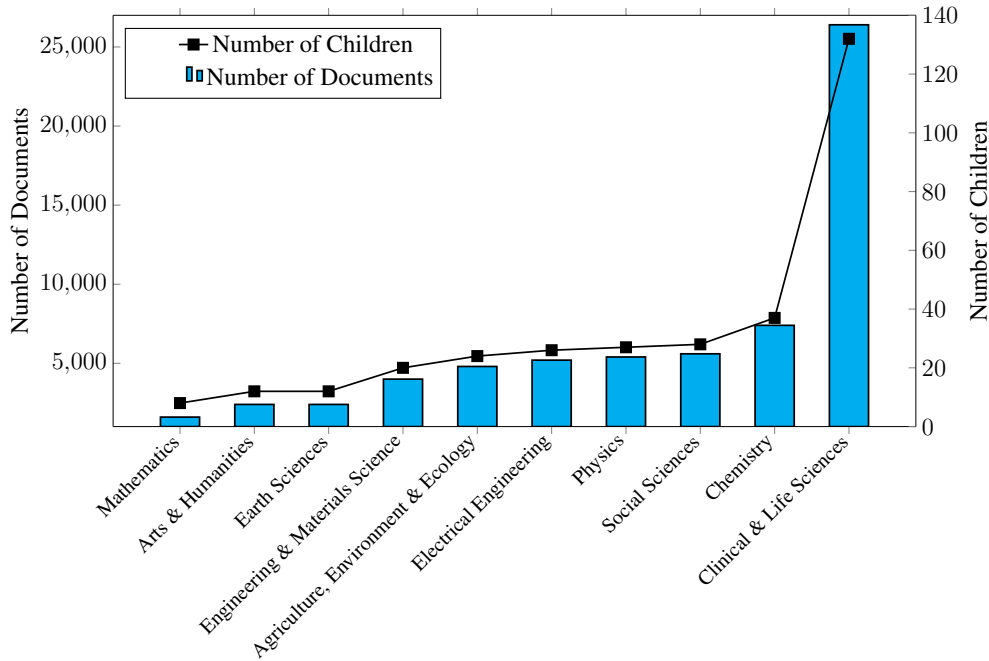
\begin{figure}[h]
  \centering
    \begin{tikzpicture}[scale = 0.65, font = \Large]
  \begin{axis}[
    symbolic x coords = {Mathematics, Arts \& Humanities, Earth Sciences, Engineering \& Materials Science, {Agriculture, Environment \& Ecology}, Electrical Engineering, Physics, Social Sciences, Chemistry, Clinical \& Life Sciences},
    axis y line*=left,
    ybar,
    ylabel={Number of Documents},
    ymin = 1000, ymax = 27000,
    xlabel near ticks,
    xticklabel style={rotate=45, anchor=east, font=\large},
    xtick = data,
    ylabel near ticks,
    bar width=0.8cm,
    enlarge x limits= 0.1,
    nodes near coords align={center},
    width=18cm,
    height = 10cm,
    legend style={
      at={(0.41,0.89)},
      draw=none
    },
  ]
\addplot[draw = black, fill = cyan, line width=1pt] 
    coordinates {(Mathematics, 1600) (Arts \& Humanities, 2400) (Earth Sciences, 2400) (Engineering \& Materials Science, 4000) ({Agriculture, Environment \& Ecology}, 4800) (Electrical Engineering, 5200) (Physics, 5400) (Social Sciences, 5600) (Chemistry, 7400) (Clinical \& Life Sciences, 26400)};
\legend{Number of Documents}
\end{axis}
  
  \begin{axis}[
    symbolic x coords = {Mathematics, Arts \& Humanities, Earth Sciences, Engineering \& Materials Science, {Agriculture, Environment \& Ecology}, Electrical Engineering, Physics, Social Sciences, Chemistry, Clinical \& Life Sciences},
    ymin = 0, ymax = 140,
    hide x axis,
    axis y line*=right,
    yticklabel pos=right,
    ylabel={Number of Children},
    ylabel near ticks,
    nodes near coords align={center},
    enlarge x limits=0.1,
    width=18cm,
    height = 10cm,
    legend style={
      at={(0.385,0.97)},
      draw=none
    },
  ]
      \addplot[color = black, mark = square*, line width=1pt, mark size=3pt] coordinates {(Mathematics, 8) (Arts \& Humanities, 12) (Earth Sciences, 12) (Engineering \& Materials Science, 20) ({Agriculture, Environment \& Ecology}, 24) (Electrical Engineering, 26) (Physics, 27) (Social Sciences, 28) (Chemistry, 37) (Clinical \& Life Sciences, 132)};
      \legend{Number of Children}
  \end{axis}
  \draw[draw=black] (0.25,6.72) rectangle (6.7,8.15);
\end{tikzpicture}
\caption[Characteristics of first-level classes of the WOS$_\text{CT}$ dataset.]{Characteristics of first-level classes of the WOS$_\text{CT}$ dataset. The bar plot gives the number of documents assigned to each class (left y-axis) while the line plot shows the number of children for each class (right y-axis).}
\label{fig:CT_stats}
\end{figure}

Figure~\ref{fig:CT_stats} shows all of the CT$_{\text{L}1}$ classes of the WOS$_\text{CT}$ dataset along with the corresponding number of children classes and number of documents. It should be noted that the CT$_{\text{L}1}$ classes with a larger number of children classes also have more assigned documents. Particularly, the ``Clinical \& Life Sciences'' class has 132 children classes, so it has a much larger number of assigned documents compared to the other classes.

\subsection{WOS$_\text{JTF}$ Dataset}
The third proposed dataset combines the JT and CT classifications with the goal of assigning classifications that capture the content of the publications more accurately. As shown above, the original dataset contains 52 JT$_{\text{L}2}$ classes and 326 CT$_{\text{L}2}$. Therefore, to combine the journal-based and citation-based classifications we attempt to map each CT$_{\text{L}2}$ class to a JT$_{\text{L}2}$ class and remove categories and documents which do not have a clear mapping between the two classification frameworks. The rationale behind this methodology is to leverage the respective advantages of the two classification approaches to create document assignments that are closely linked in the citation network and are similar based on their content.

We start by obtaining the document co-occurrence counts between each CT$_{\text{L}2}$ and JT$_{\text{L}2}$ class as a matrix $\mathbf{M} \in \mathbb{Z}^{326 \times 52}$ where each row  $\mathbf{m}_i \forall i \in \{1, \ldots, 326\}$ represents the co-occurrence counts between the $i$-th CT$_{\text{L}2}$ class (CT$_{\text{L}2, i}$) and all JT$_{\text{L}2}$ classes. In other words, for all the documents that are assigned to CT$_{\text{L}2, i}$, $\mathbf{m}_i$ captures the number of documents assigned to each of the JT$_{\text{L}2}$ classes. Therefore, $m_{i,j} \forall j \in \{1, \ldots, 52\}$ is the number of times that the $j$-th JT$_{\text{L}2}$ class (JT$_{\text{L}2, j}$) is assigned to a document belonging to CT$_{\text{L}2, i}$.

We use the co-occurrence matrix to find the most relevant JT$_{\text{L}2}$ classes for each CT$_{\text{L}2, i}$ class. First, we find the highest co-occurrence count of $\mathbf{m}_i$ as $k_i = \text{max}(\mathbf{m}_i)$ and divide $k_i$ by the co-occurrence count for each JT$_{\text{L}2}$ class to obtain a ratio for each JT$_{\text{L}2}$ class as:
\begin{equation}
    \mathbf{r}_i = \left[ \frac{k_i}{m_{i,1}}, \ldots, \frac{k_i}{m_{i,52}}\right]
\end{equation}

\noindent where $r_{i,j}$ is the $j$-th element in $\mathbf{r}_i$ that represents the ratio of the highest co-occurrence count ($k_i$) to JT$_{\text{L}2, j}$. In other words, it represents the number of times that  $k_i$ is greater than the count of JT$_{\text{L}2, j}$. We use these ratios to create a set of JT$_{\text{L}2}$ classes to which CT$_{\text{L}2, i}$ is mapped, controlled by a threshold ($\gamma$):
\begin{equation}
    Q_i = \left\{\text{JT$_{2, j}$} \forall j \in \{1, \ldots, 52\} | r_{i,j} \leq \gamma\right\}
\end{equation}

\noindent We choose a threshold of $\gamma = 1.5$, such that a CT$_{\text{L}2}$ class is only mapped to a JT$_{\text{L}2}$ class if the highest co-occurrence count for the particular CT$_{\text{L}2}$ class is less than 1.5 times the co-occurrence count of the two classes. Mappings which have fewer overlapping documents relative to the highest overlapping mapping are removed such that only the most common co-occurrences between the two classification schemas form part of the final mapping.

In order to guarantee that no misassignments occurred, two annotators individually checked each set $Q_i$ and removed JT$_{\text{L}2}$ classes from the set which are clearly inaccurate mappings from CT$_{\text{L}2, i}$ based on the class labels. We calculate the Cohen's kappa score to determine the inter-annotator agreement as:
\begin{equation}
    kappa = \frac{P_o - P_e}{1 - P_e}
\end{equation}
\noindent where $P_o$ is the proportion of items that the annotators agree on and $P_e$ is the expected agreement when annotators assign labels randomly based on empirical priors. The Cohen's kappa score obtained by the two annotators for the removal of mappings was 0.85 which indicates that there was a very strong agreement between the decisions made by the two annotators. After the initial annotation, the two annotators discussed their decision differences and agreed on the final set of mappings to remove. Table~\ref{tab:MappingRemovalExamples} lists the 10 inaccurate mappings that were removed. We use the filtered $Q_i$ set and remove 24\,215 documents that belong to a CT$_{\text{L}2}$ class with an empty mapping set $Q_i = \emptyset$. Furthermore, we remove all documents that belong to CT$_{\text{L}2, i}$ but have a JT$_{\text{L}2}$ class that is not in $Q_i$ which results in 825\,529 document removals. The reasoning behind this decision is to improve the quality of class assignments by only allowing documents which form part of the clear mapping from CT$_{\text{L}2}$ to JT$_{\text{L}2}$ classes to remain in the dataset.

\begin{table}[h]
    \caption{Removed mappings from CT$_{\text{L}2}$ to JT$_{\text{L}2}$.}
    \label{tab:MappingRemovalExamples}
    \centering
    \begin{tabular*}{\textwidth}{ccc}
        \toprule
        \normalsize CT$_{\text{L}2}$ & \normalsize JT$_{\text{L}2}$ \\
        \midrule
        Soil Science & Other social sciences \\
        Diarrhoeal Diseases & Other social sciences \\
        Water Treatment & Other social sciences \\
        Forestry & Other social sciences \\
        Nuclear Geology & Other social sciences \\
        Bioengineering & Other social sciences \\
        Folklore \& Humor & Psychology \\
        Electrical Protection & Other earth sciences \\
        Electrical - Sensors \& Monitoring & Materials sciences \\
        Remote Research \& Education & Engineering sciences (other)\\
        \bottomrule
    \end{tabular*}
\end{table}

We remove the JT$_{\text{L}2}$ classes that do not have any suitable CT$_{\text{L}2}$ classes that map to them based on the removal of mappings from the two annotators as described above. These classes include: ``Risk Assessment'', ``Operations Research \& Management Science'', ``Contamination \& Phytoremediation'', ``Herbicides, Pesticides \& Ground Poisoning'', ``Sports Science'', and ``Ornithology''. We refer to the resulting class set as JT Filtered (JTF) which comprises 6 first-level (JTF$_{\text{L}1}$) and 46 second-level (JTF$_{\text{L}2}$) classes.

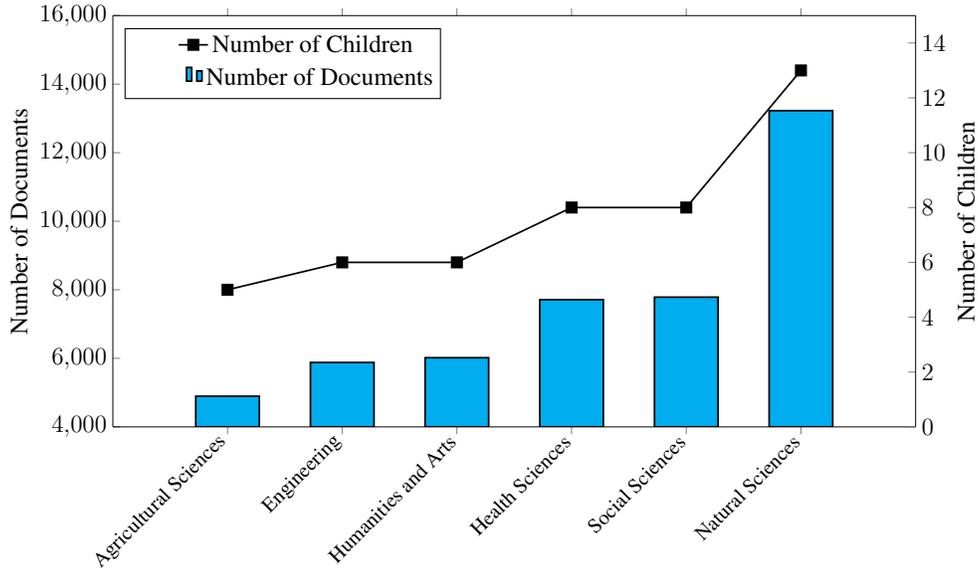
\begin{figure}[h]
    \centerline{\begin{tikzpicture}[scale = 0.65, font = \Large]
  \begin{axis}[
    symbolic x coords = {Agricultural Sciences, Engineering, Humanities and Arts, Health Sciences, Social Sciences, Natural Sciences},
    axis y line*=left,
    ybar,
    ylabel={Number of Documents},
    ymin = 4000, ymax = 16000,
    xlabel near ticks,
    xticklabel style={rotate=45, anchor=east, font=\large},
    xtick = data,
    ylabel near ticks,
    bar width=1.3cm,
    enlarge x limits= 0.2,
    nodes near coords align={center},
    width=18cm,
    height=10cm,
    legend style={
      at={(0.41,0.89)},
      draw=none
    },
  ]
\addplot[draw = black, fill = cyan, line width=1pt] 
    coordinates {(Agricultural Sciences, 4895) (Engineering, 5879) (Humanities and Arts, 6018) (Health Sciences, 7710)  (Social Sciences, 7785) (Natural Sciences, 13226)};
\legend{Number of Documents}
\end{axis}
  
  \begin{axis}[
    symbolic x coords = {Agricultural Sciences, Engineering, Humanities and Arts, Health Sciences, Social Sciences, Natural Sciences},
    ymin = 0, ymax = 15,
    hide x axis,
    axis y line*=right,
    yticklabel pos=right,
    ylabel={Number of Children},
    ylabel near ticks,
    nodes near coords align={center},
    enlarge x limits=0.2,
    width=18cm,
    height=10cm,
    legend style={
      at={(0.385,0.97)},
      draw=none
    },
  ]
      \addplot[color = black, mark = square*, line width=1pt, mark size=3pt] coordinates {(Agricultural Sciences, 5) (Engineering, 6) (Humanities and Arts, 6) (Health Sciences, 8)  (Social Sciences, 8) (Natural Sciences, 13)};
    \legend{Number of Children}
  \end{axis}
  \draw[draw=black] (0.25,6.72) rectangle (6.7,8.15);
  
\end{tikzpicture}}
\caption[Characteristics of first-level classes of the WOS$_\text{JTF}$ dataset.]{Characteristics of first-level classes of the WOS$_\text{JTF}$ dataset. The bar plot gives the number of documents assigned to each class (left y-axis) while the line plot shows the number of children for each class (right y-axis).}
\label{fig:HCRD_stats}
\end{figure}

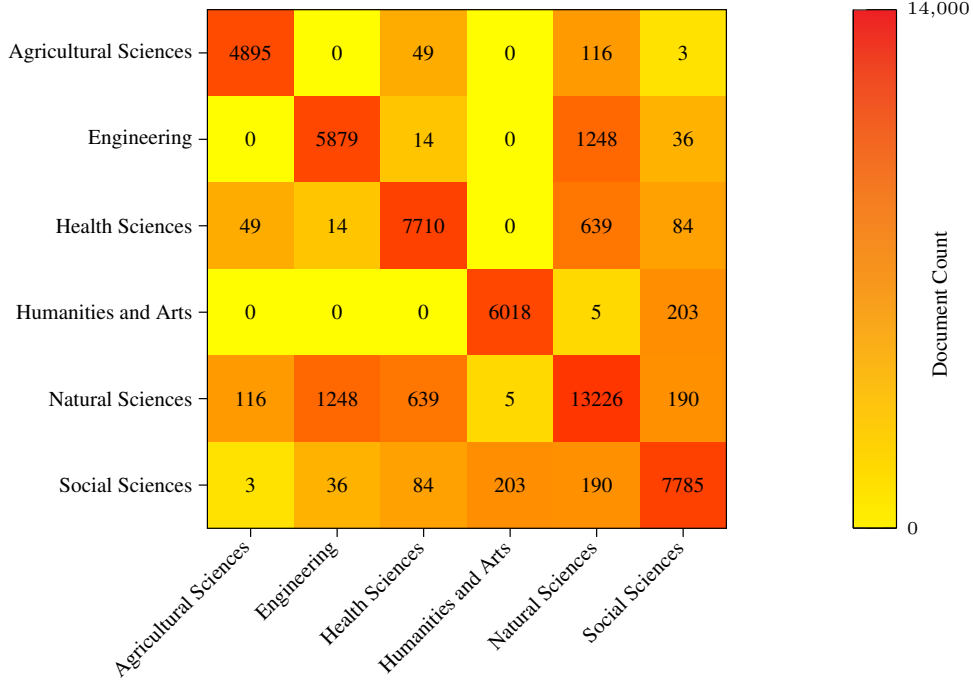
\begin{figure}[h]
  \centerline{\scalebox{1.15}{\begin{tikzpicture}

    \definecolor{lightred}{RGB}{255, 50, 0}

    \pgfplotsset{
        colormap={mypurpleyellow}{
            color(0cm) = (yellow);
            color(6cm) = (red);
        }
    }

    \def\matrixdata{{4895,0,49,0,116,3},
    {0,5879,14,0,1248,36},
    {49,14,7710,0,639,84},
    {0,0,0,6018,5,203},
    {116,1248,639,5,13226,190},
    {3,36,84,203,190,7785}}
    
    \def\rowlabels{{"Agricultural Sciences", "Engineering", "Health Sciences", "Humanities and Arts", "Natural Sciences", "Social Sciences"}}
    
    \foreach \row [count=\y] in \matrixdata {
        \foreach \cell [count=\x] in \row {
            \pgfmathsetmacro\normalizedvalue{100*(ln(\cell+1)/ln(16000))}
            \pgfmathsetmacro\normalizedvalue{max(\normalizedvalue, 1)} 
            \node[fill=lightred!\normalizedvalue!yellow, text=black, minimum size=1cm, inner sep=0pt, anchor=center, font=\scriptsize] at (\x, -\y) {\cell};
        }
    }
    
    \foreach \y/\label in {1/Agricultural Sciences, 2/Engineering, 3/Health Sciences, 4/Humanities and Arts, 5/Natural Sciences, 6/Social Sciences} {
        \node[anchor=east, font=\scriptsize] at (0.45, -\y) {\label};
        \draw (0.4, -\y) -- (0.5, -\y);
    }

    \foreach \x/\label in {1/Agricultural Sciences, 2/Engineering, 3/Health Sciences, 4/Humanities and Arts, 5/Natural Sciences, 6/Social Sciences} {
        \node[anchor=east, font=\scriptsize, rotate=45] at (\x+0.05, -6.6) {\label};
        \draw (\x, -6.5) -- (\x, -6.6);
    }

    \draw[black, thin] (0.5, -0.5) rectangle (6.5, -6.5);

    \begin{axis}[
        at={(800,-130)},  
        anchor=south west,
        scale only axis,
        width=1cm,
        height=6cm,  
        colorbar,  
        hide axis,  
        point meta min=0,
        point meta max=14000,  
        colorbar style={
            ytick={0, 14000},  
            yticklabel style={font=\tiny},
            yticklabel pos=right,  
            colormap name=mypurpleyellow,  
            ylabel style={font=\scriptsize, at={(2, 0.27)}, anchor=west},
            ylabel={Document Count},
            scaled ticks=false
        },
    ]
    \end{axis}
    
\end{tikzpicture}}}
    \caption{Co-occurrence counts for first-level classes of the WOS$_\text{JTF}$ dataset.}
\label{fig:HCRD_overlap_level_1}
\end{figure}

To create the WOS$_\text{JTF}$ dataset, we sample 1\,000 documents for each of the JTF$_{\text{L}2}$ classes to obtain a total of 42\,926 instances. Figure~\ref{fig:HCRD_stats} presents the first-level classes of the WOS$_\text{JTF}$ dataset with the associated number of children classes and number of documents assigned to each class while Figure~\ref{fig:HCRD_overlap_level_1} presents the co-occurrence counts for the first-level classes. From Figure~\ref{fig:HCRD_overlap_level_1} we can see that there is a much clearer distinction between the first-level class assignments in the WOS$_\text{JTF}$ dataset compared to the WOS$_\text{JT}$ dataset, with fewer instances overlapping multiple first-level classes. 

\begin{figure}[h]
\centerline{\begin{tikzpicture}[scale = 0.65, font = \Large]
        \begin{axis}[
            ybar,
            xticklabel=\pgfmathprintnumber\tick,
            xlabel={Documents per Class},
            ylabel={Frequency},
            ymin=0,
            bar width=0.2cm,
            enlarge x limits=0.02,
            nodes near coords align={center},
            width=20cm,
            height=10cm,
            xmin=250,
            xmax=2250,
            xtick={250,500,750,1000,1250,1500,1750,2000,2250},
            xticklabels={250,500,750,1000,1250,1500,1750,2000,2250},
        ]
        
        \addplot+[hist={bins=32}, fill=cyan, draw=black, line width=1pt]
        table[y index=0] {HCRD_lower_class_count.txt};
        \end{axis}

        \begin{axis}[
        hide axis,
        at={(0,0)},
        width=20cm,
        height=10cm,
        xmin=250,
        xmax=2250,
        enlarge x limits=0.02,
    ]
    
    \addplot[color=blue, mark=none, line width=1.2pt] table {HCRD_kde_data.txt};
    
    \end{axis}
        
\end{tikzpicture}}
    \caption{Distribution of the number of documents assigned to each of the JTF$_{\text{L}2}$ classes in the WOS$_\text{JTF}$ dataset.}
\label{fig:HCRD_label_counts_level_2}
\end{figure}
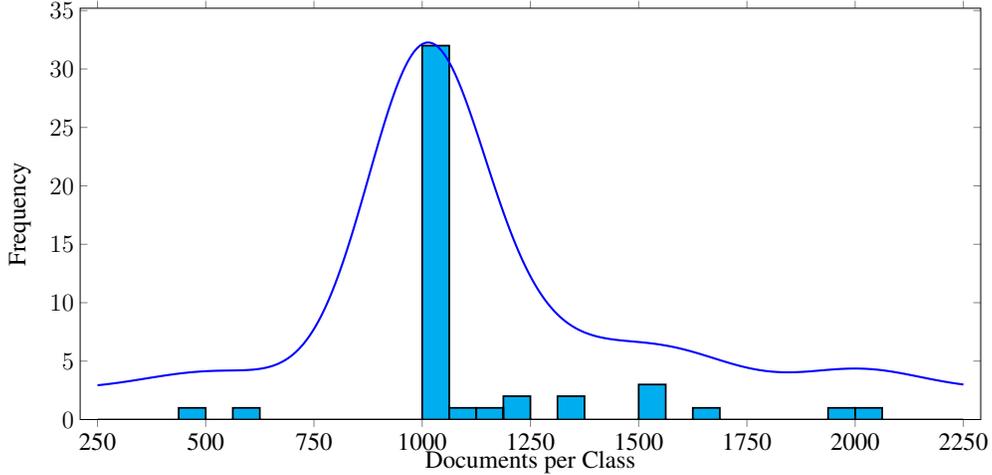

Figure~\ref{fig:HCRD_label_counts_level_2} gives the distribution of documents assigned to the second-level classes of WOS$_\text{JTF}$. This figure shows that most classes have around 1\,000 document assignments. However, ``Biochemistry \& molecular biology'' and ``Marine \& freshwater biology'' only have 450 and 594 documents respectively since they do not have 1\,000 documents to sample from. Furthermore, ``Clinical and public health (other)'' and ``Engineering sciences (other)'' have 1\,979 and 2\,045 documents respectively since they often co-occur with other classes.

\section{Evaluation}
\subsection{Cluster Analysis}
We evaluate the quality of the three datasets by analysing the semantic similarity between the documents belonging to each of the classes (or clusters). We investigate the semantic similarity between the documents within each class-cluster as well as the separation between documents in different clusters to determine whether the class assignments form cohesive clusters of documents that are closely related thematically. We evaluate the class assignments separately for the two levels of the class hierarchy.

First, we use a sentence-BERT model~\citep{Reimers2019SentenceBERTSE} to convert the title and abstract of each document in the dataset to a fixed-sized embedding which captures the semantic meaning of the document. A sentence-BERT model is a modified BERT model that is tuned to produce a semantic embedding for a sentence or document in contrast to the standard BERT model which produces embeddings for each token in the input sequence. We use the \texttt{all-mpnet-base-v2} sentence-BERT model from the SentenceTransformers library. Furthermore, we cluster (or group) all of the documents based on their first- or second-level class assignments. Documents that belong to more than one class at a level are duplicated and placed into each of the class-clusters they belong to.
    
Suppose we have $L$ clusters for a particular level of a dataset. We use the document embeddings to calculate the average cosine similarity between the instances in each cluster $j \in \{1, \ldots, L\}$ as:
\begin{equation}
    o_j = \frac{1}{N_j(N_j-1)} \sum_{k=1}^{N_j} \sum_{l=1, k \neq l}^{N_j} \text{CosSim}(\mathbf{x}_k, \mathbf{x}_l)
\end{equation}
\noindent where $N_j$ is the number of instances that belong to cluster $j$ and CosSim is the cosine similarity function. The cosine similarity function between the embeddings of two documents is calculated as:
\begin{equation}
    \text{CosSim}(\mathbf{x}_k, \mathbf{x}_l) =\frac{\mathbf{x}_k \cdot \mathbf{x}_l}{\|\mathbf{x}_k\| \cdot \|\mathbf{x}_l\|}
\end{equation}

\noindent such that semantically similar documents have high similarity scores.

To determine how well-separated the documents in the different clusters are, we calculate the silhouette score for each instance $\mathbf{x}_i \forall i \in \{1, \ldots, N\}$, where $N$ is the number of instances in the dataset. The silhouette score measures the quality of clusters by quantifying how well-separated and internally cohesive the clusters are. The silhouette score for instance $i$ is calculated as:
\begin{equation}
    s_i = \frac{b(i) - a(i)}{\max(a(i), b(i))}
\end{equation}

\noindent where $a(i)$ is the average distance from instance $i$ to all other instances in the same cluster which we calculate with the cosine distance function given by:
    \begin{equation}
        \text{CosDist}(\mathbf{x}_k, \mathbf{x}_l) = 1 - \text{CosSim}(\mathbf{x}_k, \mathbf{x}_l)
    \end{equation}
    \noindent and $b(i)$ is the average distance from instance $i$ to all instances in the nearest neighbouring cluster. The nearest neighbouring cluster is determined by minimising the average distance from the instance to the instances of a different cluster. The silhouette score ranges from -1 to 1 where a high silhouette score (1) indicates that instances are semantically similar within a cluster and are far apart from other clusters while a low silhouette score (-1) indicates that instances are not well-clustered and the clusters have a high overlap.

\begin{figure}[h]
  \centering
  \begin{subfigure}{1\textwidth}
      \includegraphics[width=1\linewidth]{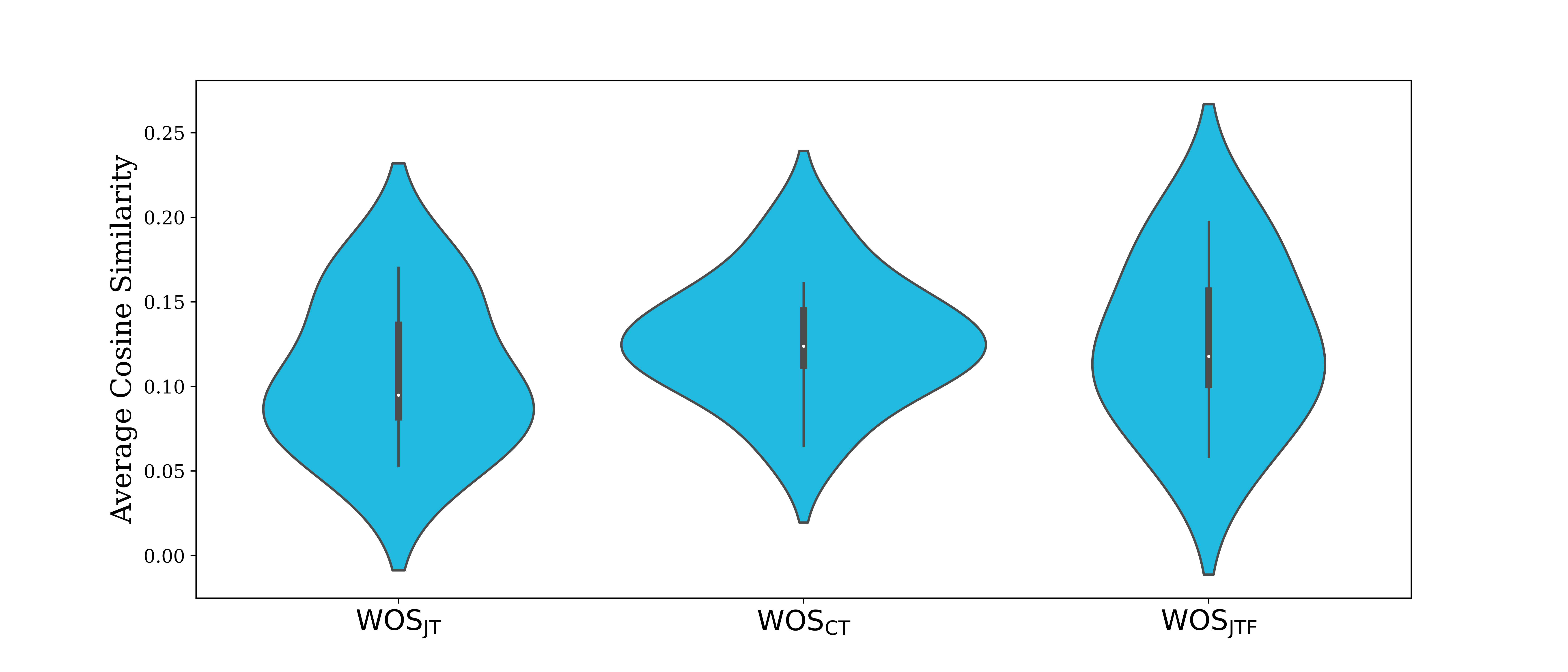}
      \vspace{-5ex}
      \caption{First-level classes.}
  \end{subfigure}
\begin{subfigure}{1\textwidth}
    \includegraphics[width=1\linewidth]{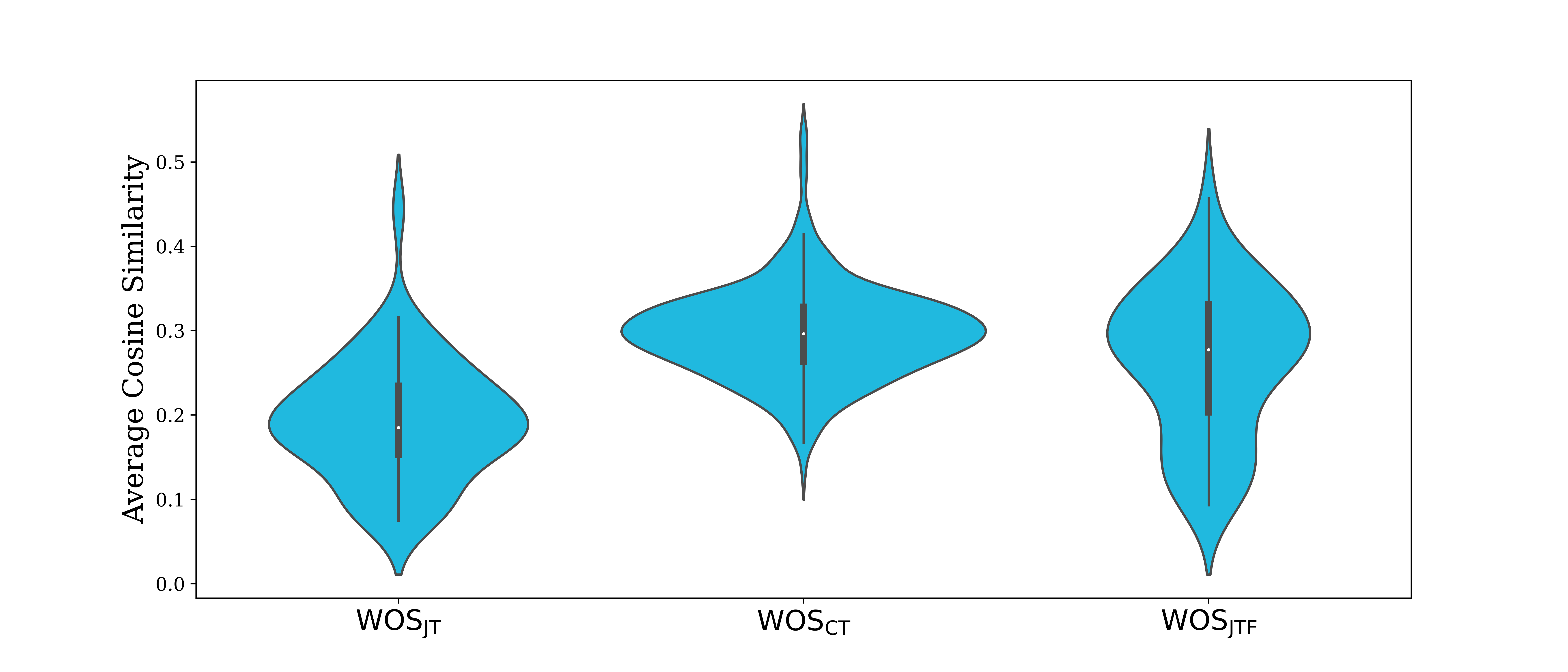}
    \vspace{-5ex}
    \caption{Second-level classes.}
\end{subfigure}
\caption{Violin plots of the cosine similarity per class-cluster for the three newly proposed datasets.}
\label{fig:combined_cluster_similarity_dist}
\end{figure}

The violin plots in Figure~\ref{fig:combined_cluster_similarity_dist} gives the cosine similarity distributions over the classes in the first and second level for the three datasets. These plots show that the semantic similarity between documents in the same classes is generally higher in the WOS$_\text{JTF}$ dataset than the WOS$_\text{JT}$ dataset. This shows that the proposed filtering approach was able to leverage the citation-based classifications to improve the classification of the documents such that the documents belonging to a certain class are on average more semantically similar than the WOS$_\text{JT}$ assignments.

\begin{figure}[h]
  \centering
  \begin{subfigure}{1\textwidth}
      \includegraphics[width=1\linewidth]{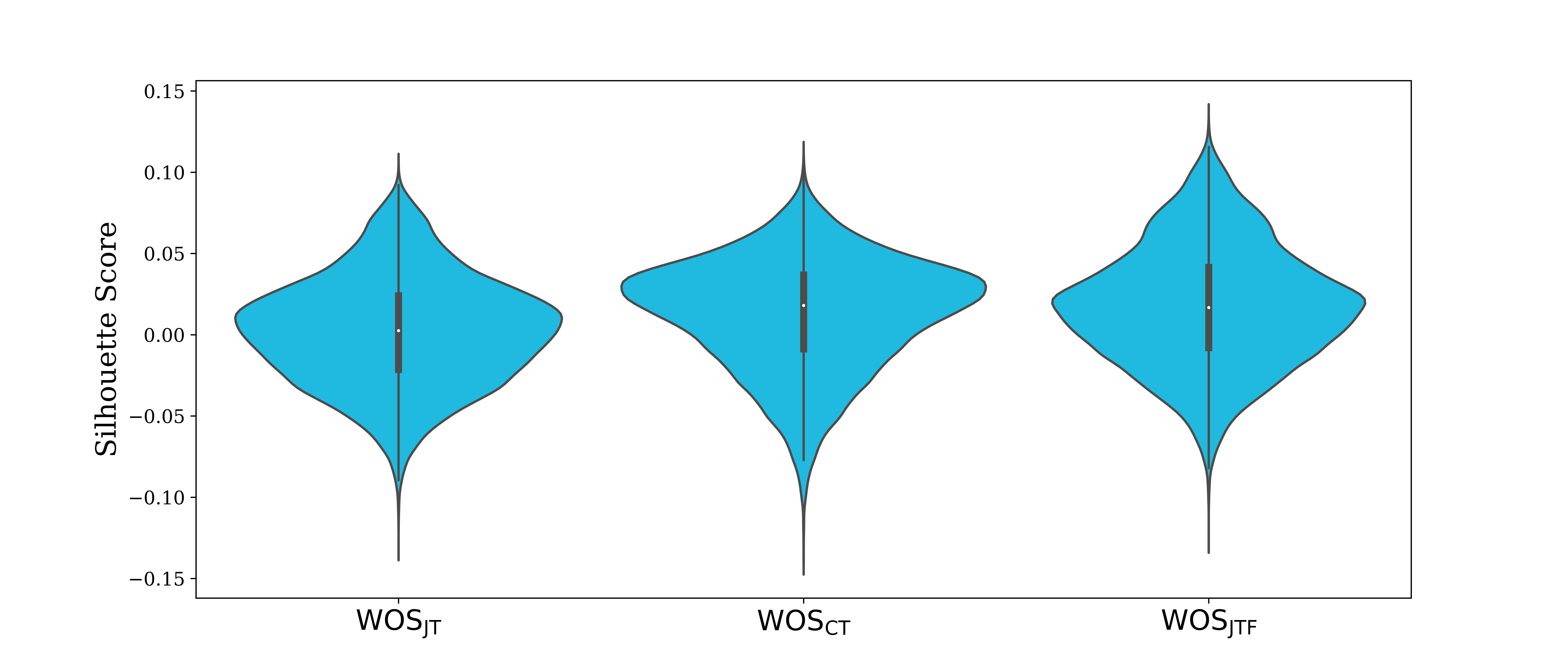}
      \vspace{-5ex}
      \caption{First-level classes.}
  \end{subfigure}
\begin{subfigure}{1\textwidth}
    \includegraphics[width=1\linewidth]{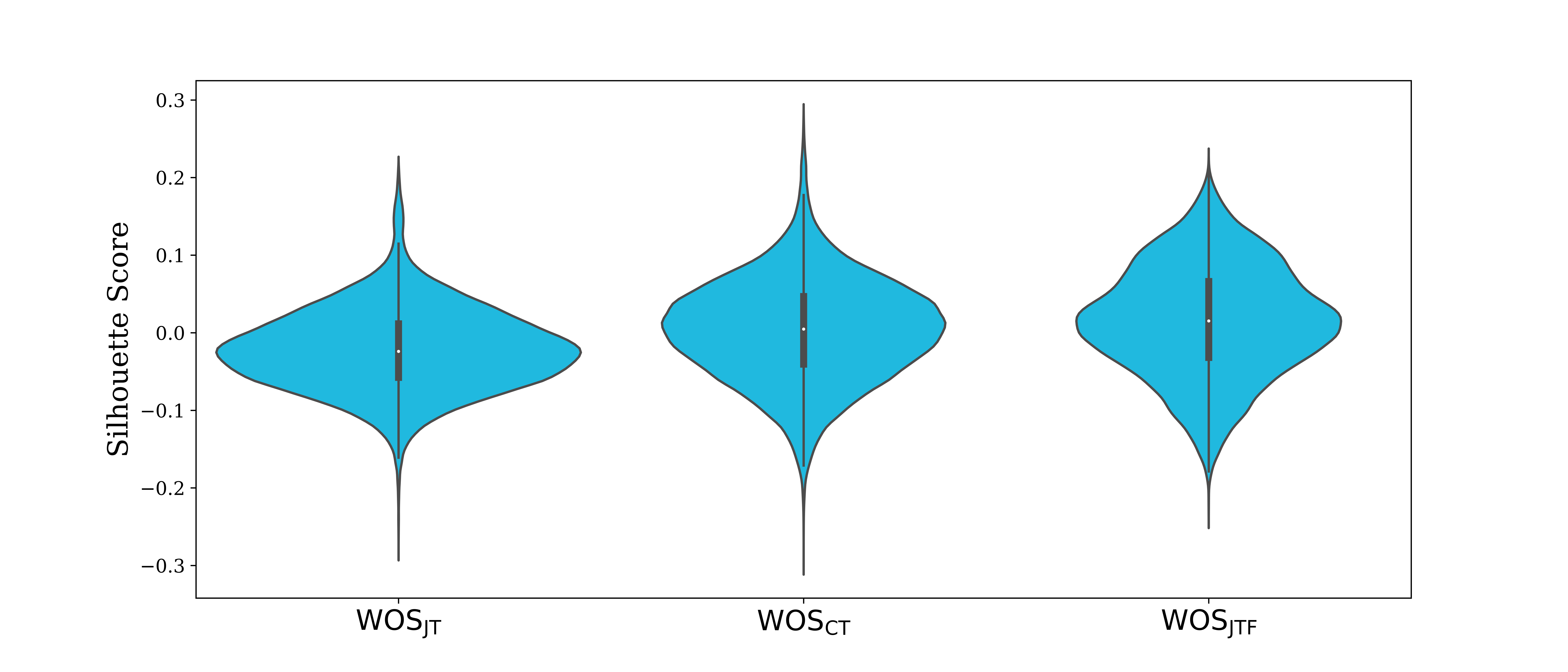}
    \vspace{-5ex}
    \caption{Second-level classes.}
\end{subfigure}
\caption{Violin plots of the silhouette scores for the three newly proposed datasets.}
\label{fig:combined_silhouette_dist}
\end{figure}

Figure~\ref{fig:combined_silhouette_dist} presents the violin plots for the silhouette scores over all of the instances for the first and second level classes on the three datasets. The figures show that the WOS$_\text{JTF}$ dataset generally has a higher silhouette score than the WOS$_\text{JT}$ dataset, especially for the second level of the class hierarchy. This indicates that our approach to create the WOS$_\text{JTF}$ dataset was able to effectively remove instances and categories based on the citation-based classifications in order to improve the separation in terms of semantic similarity between the class-clusters on average. However, the silhouette scores are still very low across the three datasets which implies that many of the clusters are not well-separated and may overlap.

\subsection{Classification Results}
We perform experiments on the three newly created datasets with four state-of-the-art HTC approaches: HPTD-ELECTRA~\citep{dutoit2}, HPTD-DeBERTaV3~\citep{dutoit2}, GHLA$_\text{RoBERTa}$~\citep{dutoit}, and HPT~\citep{Wang2}. For each approach we use the same architecture and hyperparameter tuning procedure as mentioned in the original papers.


Table~\ref{tab:StabilityResults} presents the results of the different approaches on the three newly created datasets with standard deviations over three runs with different seeds given in parentheses. The results show that GHLA$_\text{RoBERTa}$ and HPTD-DeBERTaV3 generally outperform the other approaches and they obtain very similar results for each of the datasets. The largest performance difference between GHLA$_\text{RoBERTa}$ and HPTD-DeBERTaV3 is an improvement of 0.19 for the WOS$_\text{JT}$ Macro-F1 score, showing that they perform consistently similar on each of the datasets. Furthermore, the HPT approach obtains the highest Macro-F1 score on the WOS$_\text{CT}$ dataset and outperforms HPTD-ELECTRA on all three datasets.

The results show that the classification performance is significantly improved on the WOS$_\text{JTF}$ dataset compared to WOS$_\text{JT}$. For example, the Micro-F1 and Macro-F1 scores of the GHLA$_\text{RoBERTa}$ approach increase by 17.34 and 20.54 respectively when moving from WOS$_\text{JT}$ to WOS$_\text{JTF}$. We believe that this increase in performance is due to the filtering approaches used to remove documents and categories based on the mapping from the CT classifications. This indicates that the WOS$_\text{JTF}$ dataset contains more accurate classifications (or fewer incorrectly assigned documents) which can be more effectively learnt by the classification models. However, other factors such as the number of classes and the average classes per instance may also explain a proportion of the observed performance differences.

In terms of model classification consistency, the GHLA$_\text{RoBERTa}$ and HPT approaches generally have the most stable results over multiple runs with GHLA$_\text{RoBERTa}$ obtaining the lowest standard deviations on the WOS$_\text{JTF}$ and WOS$_\text{JT}$ datasets and HPT having the lowest standard deviation on the WOS$_\text{CT}$ dataset.


\begin{table}[h]
    \caption[The average performance results of evaluating the models over three independent runs on the newly proposed datasets.]{The average performance results of evaluating the models over three independent runs on the newly proposed datasets. The values in parentheses show the corresponding standard deviations.}
    \centering
    \label{tab:StabilityResults}
    \begin{tabular*}{\textwidth}{lcccccc}
        \toprule
        \normalsize Model &
        \multicolumn{2}{c}{\normalsize WOS$_\text{JTF}$} &
        \multicolumn{2}{c}{\normalsize WOS$_\text{JT}$} &
        \multicolumn{2}{c}{\normalsize WOS$_\text{CT}$} \\
        \cmidrule(lr){2-3}\cmidrule(lr){4-5}\cmidrule(lr){6-7}
        & {Micro-F1} & {Macro-F1} & {Micro-F1} & {Macro-F1} & {Micro-F1} & {Macro-F1} \\
        \midrule
        HPT & 84.97 (0.17) & 82.13 (0.18) & 67.62 (0.18) & 61.71 (0.17) & 73.25 (\textbf{0.12}) & \bf{61.87} (\textbf{0.15}) \\
        HPTD-ELECTRA & 84.75 (0.38) & 81.70 (0.42) & 67.19 (0.20) & 60.91 (0.26) & 71.39 (0.39) & 58.41 (0.68) \\
        HPTD-DeBERTaV3 & 85.68 (0.14) & \bf{82.93} (0.15) & 68.35 (0.12) & 62.19 (0.46) & \bf{73.45} (0.14) & 61.27 (0.56) \\
        GHLA$_\text{RoBERTa}$ & \bf{85.72} (\textbf{0.11}) & 82.92 (\textbf{0.13}) & \bf{68.38} (\textbf{0.11}) & \bf{62.38} (\textbf{0.11}) & 73.34 (0.25) & 61.29 (0.40) \\
        \bottomrule
    \end{tabular*}
\end{table}

\section{Conclusion}
In this paper we introduced three new benchmark Hierarchical Text Classification (HTC) datasets in the domain of research publications comprising papers' titles and abstracts collected from the Web of Science publication database. Using this data, we created two HTC datasets that are derived from existing journal-and citation-based classification schemas respectively. These schemas have different disadvantages such as the inaccurate journal-based classifications due to journals that cover research topics that are too broad and the citation-based classifications that only allow a document to belong to a single research field. Therefore, we proposed an approach to combine these two classifications to increase the probability of a document being assigned to the correct classes while allowing documents to be classified into more than one research field. To create the datasets, we sampled documents equally for each of the second-level classes such that our datasets are significantly more balanced than other HTC benchmark datasets. We evaluated the quality of the proposed datasets through a clustering-based analysis and showed that our proposed approach which combines the two classification schemas resulted in documents with the same classes being semantically more similar. Finally, we performed experiments on the three datasets with four state-of-the-art HTC approaches to provide a baseline for future research.

\bibliographystyle{unsrtnat}
\bibliography{references}  






\end{document}